\RequirePackage{ifpdf}
\ifpdf 
\documentclass[pdftex]{sigma}
\else
\documentclass{sigma}
\fi

\begin{document}

\allowdisplaybreaks

\renewcommand{\thefootnote}{$\star$}

\renewcommand{\PaperNumber}{029}

\FirstPageHeading

\ShortArticleName{Supersymmetry Transformations for Delta Potentials}

\ArticleName{Supersymmetry Transformations for Delta Potentials\footnote{This
paper is a contribution to the Proceedings of the Workshop ``Supersymmetric Quantum Mechanics and Spectral Design'' (July 18--30, 2010, Benasque, Spain). The full collection
is available at
\href{http://www.emis.de/journals/SIGMA/SUSYQM2010.html}{http://www.emis.de/journals/SIGMA/SUSYQM2010.html}}}

\Author{David J. FERN\'ANDEZ C.~$^\dag$, Manuel GADELLA~$^\ddag$ and Luis Miguel NIETO~$^\ddag$}

\AuthorNameForHeading{D.J.~Fern\'andez C., M.~Gadella and L.M. Nieto}

\Address{$^\dag$~Departamento de F\'isica, Cinvestav, AP 14-740, 07000 M\'exico DF, Mexico}
\EmailD{\href{mailto:david@fis.cinvestav.mx}{david@fis.cinvestav.mx}}

\Address{$^\ddag$~Departamento de F\'isica Te\'orica, At\'omica y
Optica, Facultad de Ciencias,\\
\hphantom{$^\ddag$}~47041 Valladolid, Spain}
\EmailD{\href{mailto:manuelgadella1@gmail.com}{manuelgadella1@gmail.com}, \href{mailto:luismi@metodos.fam.cie.uva.es}{luismi@metodos.fam.cie.uva.es}}

\ArticleDates{Received November 30, 2010, in f\/inal form March 19, 2011;  Published online March 22, 2011}

\Abstract{We make a detailed study of the f\/irst and second-order
SUSY partners of a one-dimensional free Hamiltonian with a singular
perturbation proportional to a Dirac delta function. It is shown
that the second-order transformations increase the spectral
manipulation possibilities of\/fered by the standard f\/irst-order
supersymmetric quantum mechanics.}


\Classification{81Q60}

\renewcommand{\thefootnote}{\arabic{footnote}}
\setcounter{footnote}{0}

\section{Introduction}

The study of one-dimensional Hamiltonians with a point interaction
has received renewed attention during the past two decades
\cite{S,K,C,C1,H,H1,Z}. In general, a point interaction is described
by a potential concentrated either in a single or a discrete number
of points as it happens, e.g., for the Dirac delta or its
derivative. Mathematically, in order to def\/ine these potentials, we
use the theory of extensions of symmetric operators with equal
def\/iciency indices. These extensions have domains which are
characterized by some matching conditions for the wave functions at
the points supporting the interaction  \cite{K,A,AK,GKN,GNN}. In
particular, the Dirac delta barrier or well have been extensively
studied in this way with or without other interactions \cite{FA},
with or without mass discontinuities at the singular points etc.~\cite{GHNN,AGHN}.

On the other hand, supersymmetric quantum mechanics (SUSY QM) has
emerged as the standard technique for generating new potentials with
known spectra departing from an initial one
\cite{wi81,mi84,fe84,abi84a,abi84b,su85a,su85b,su86,su87,bdh87,ad88,lrb90,rrr91,lr91,ba93,sb95,ba01,cks01,mr04,ac04,pl04,su05,ff05,do07,as07,so08,fe10}.
The method has been applied successfully to regular one-dimensional
potentials def\/ined on the full real line \cite{fgn98,fhm98}, on the
positive semi-axis~\cite{ro98a,ro98b} or in a f\/inite
interval~\cite{cf08}. Although there are some works dealing with SUSY
QM applied to point potentials
\cite{dnnr99,ut03a,ut03b,ftc03,pl1,pl2,pl3}, however the
corresponding study has been done just for particular f\/irst-order
SUSY transformations, without analyzing the full possibilities of
spectral manipulation of\/fered by the method. It is interesting to
note as well that a point potential may appear as hidden
supersymmetries~\cite{pl4,pl5}.

Now, it is the appropriate time for studying the behavior of point
potentials with bound states under SUSY QM. Due to the calculation
complexity, we shall focus our attention to f\/irst and second-order
transformations, which anyway are interesting by themselves
\cite{ais93,aicd95,bs97,fe97}. We shall restrict the discussion to
the following one-dimensional Hamiltonian
\begin{gather}\label{1}
    H_0=-\frac12\,\frac{d^2}{dx^2}+V_0(x) ,\qquad V_0(x):=
    -a\delta(x) ,\qquad a>0 ,
\end{gather}
which is mathematically well def\/ined and self-adjoint provided that
we use as its domain $\cal D$ the subspace of the Sobolev space
$W^2_2({\mathbb R}/\{0\})$ such that for any $\psi(x)\in\cal D$, one
has:
\begin{gather}\label{2}
\left(
 \begin{array}{c}
    \psi(0+) \\[2ex]
 \psi'(0+)
               \end{array}
             \right)  =  \left(
      \begin{array}{cc}
        1 & 0 \\[2ex]
        -2a & 1 \\
      \end{array}
    \right) \left(
               \begin{array}{c}
    \psi(0-) \\[2ex]
 \psi'(0-) \\
               \end{array}
             \right) ,
\end{gather}
where $\psi(0+)$, $\psi'(0+)$ and $\psi(0-)$, $\psi'(0-)$ are the
right and left limits of $\psi(x)$, $\psi'(x)$ at the origin
respectively~\cite{K}.

In order to achieve our goal, we have organized this paper as
follows: in Section~\ref{section2}, we will study the solutions of the stationary
Schr\"odinger equation for the Hamiltonian $H_0$ given by~(\ref{1}).
In Section~\ref{section3} we will apply the f\/irst-order SUSY techniques, in
Section~\ref{section4} we will analyse the second-order transformations and in
Section~\ref{section5} we will present our conclusions.

\section{Solution of the Schr\"odinger equation}\label{section2}

Let us evaluate in the f\/irst place the general solution of the
stationary Schr\"odinger equation for an arbitrary $\epsilon =
-k^2/2 < 0$:
\begin{gather}\label{se}
H_0 u(x)=\epsilon u(x) ,
\end{gather}
with $H_0$ given in (\ref{1}). There is one solution vanishing for
$x\rightarrow -\infty$, denoted $u_+(x)$, of the form
\begin{gather}\label{7}
u_+(x)=e^{kx} H(-x)+(\alpha\,e^{kx}+\beta e^{-kx})H(x) , \qquad
k>0 ,
\end{gather}
where $H(x)$ is the Heaviside step function, and $\alpha$, $\beta$
are constants to be determined from the discontinuity equations \eqref{2}.
We need as well the derivative of $u_+(x)$,
\begin{gather}\label{du}
u'_+(x) = k u_+(x) - 2k\beta e^{-kx}H(x) + (\alpha + \beta - 1)
\delta(x).
\end{gather}
From equations (\ref{7}) and (\ref{du}) it turns out that
\begin{gather}\label{9}
  u_+(0+)=\alpha+\beta ,\qquad u_+(0-)=1 ,\qquad
u_+'(0+)=k(\alpha-\beta), \qquad u_+'(0-)=k.
\end{gather}
On the other hand, using equations (\ref{2}) and (\ref{9}), we
obtain:
\begin{gather*}
\alpha+\beta = 1  , \qquad \alpha-\beta = 1 - 2\tilde a  ,
\end{gather*}
where $\tilde a = a/k$. Hence:
\begin{gather*}
\alpha = 1 - \tilde a ,\qquad \beta = \tilde a  .
\end{gather*}
Inserting these expressions in equations (\ref{7}) and (\ref{du}),
we f\/inally get
\begin{gather}
  \label{up}
u_+(x)=e^{kx} -\tilde a \big( e^{kx} - e^{-kx}\big) H(x) , \\
  u'_+(x) = k u_+(x) - 2 a e^{-kx}H(x) .\nonumber
\end{gather}

Note that the Hamiltonian $H_0$ in equation (\ref{1}) is invariant
under the change $x\rightarrow -x$. Thus, we can f\/ind a second
linearly independent solution $u_-(x)$ for the same $\epsilon =
-k^2/2$, vanishing now for $x \rightarrow \infty$, by applying this
transformation to $u_+(x)$:
\begin{gather}\label{um}
u_-(x) = \tilde a \big( e^{kx} - e^{-kx} \big)H(-x) + e^{- kx} .
\end{gather}
Moreover:
\begin{gather*}
u'_-(x) = - k u_-(x) + 2 a e^{kx}H(-x) .
\end{gather*}
Finally, the general solution of equation (\ref{se}) for $\epsilon =
-k^2/2 < 0$ is a linear combination of both~(\ref{up}) and~(\ref{um}) which, up to an unessential constant factor, becomes:
\begin{gather}\label{22}
u(x)  =  u_+(x) + D  u_-(x)
 = e^{kx} + D   e^{- kx} - \tilde a \big(e^{kx} - e^{-kx}\big)
\left[ H(x) - D H(-x)\right]  ,
\end{gather}
where $D$ is a constant. The corresponding derivative is given by:
\begin{gather}
u'(x)   =   - k u(x) + 2 k e^{k x}\left[ 1 -\tilde a H(x) + D\tilde
a H(-x) \right]  . \label{upgenb0}
\end{gather}

Note that, up to normalization, both solutions $u_\pm(x)$ lead to
the same bound state for $k_0 = a$:
\begin{gather*}
\psi_0(x) = \sqrt{a} \big[e^{k_0 x} H(-x) + e^{-k_0
x} H(x)\big].
\end{gather*}
The corresponding eigenvalue becomes
\begin{gather*}
E_0 = - \frac{a^2}{2},
\end{gather*}
which coincides with the result derived in \cite{GNN}.

On the other hand, the scattering states for $\epsilon = \kappa^2/2
> 0$ can be simply obtained from the solutions given in equations
(\ref{up}), (\ref{um}) by the substitution $k \rightarrow -i \kappa$, $\kappa > 0$. In particular, for a~probability f\/lux approaching the
singularity from $-\infty$ the corresponding scattering state arises
in this way from the $u_-(x)$ of equation (\ref{um}), which (up to
unessential constant factor) leads to
\begin{gather}\label{ss}
\psi(x) = \left[e^{i \kappa x} + \frac{ia}{\kappa - ia}e^{-i \kappa
x} \right] H(-x) + \frac{\kappa}{\kappa - ia}e^{i \kappa x} H(x).
\end{gather}
It is clear now that the ref\/lection $R$ and transmition $T$
coef\/f\/icients become the standard ones (see, e.g., \cite{flugge}):
\begin{gather} \label{reftracoe}
R = \left\vert \frac{ia}{\kappa - ia} \right\vert^2 =
\frac{a^2}{\kappa^2 + a^2}, \qquad T = \left\vert
\frac{\kappa}{\kappa - ia} \right\vert^2 = \frac{\kappa^2}{\kappa^2
+ a^2}.
\end{gather}

\section{First-order SUSY transformations}\label{section3}

Let us start with the initial Schr\"odinger Hamiltonian $H_0$ given
in (\ref{1}). As it is well known (see, e.g., \cite{ff05,fe10} and
the references cited there), its f\/irst-order SUSY partner,
\begin{gather*}
H_1=-\frac12 \frac{d^2}{dx^2}+V_1(x),
\end{gather*}
is intertwined with $H_0$ in the way
\begin{gather}\label{1interwining}
H_1 A_1^+ = A_1^+ H_0,
\end{gather}
where
\begin{gather*}
A_1^+ = \frac{1}{\sqrt{2}}\left(- \frac{d}{dx} +
\frac{u'}{u}\right) .
\end{gather*}
Here, the transformation function $u(x)$ is the seed solution given
in (\ref{22}), associated to the factorization energy $\epsilon = -
k^2/2$ and satisfying equation (\ref{se}). The SUSY partner
potential $V_1(x)$ of $V_0(x)$ is given by:
\begin{gather}\label{6}
V_1(x) = V_0(x) - [\ln u(x)]'' .
\end{gather}
We assume the standard restriction $\epsilon\leq E_0 \Rightarrow
k\geq k_0$, in order to avoid the creation of new singularities in
$V_1(x)$ with respect to those of $V_0(x)$. Note that, from equation
(\ref{6}) and (\ref{se}) we have:
\begin{gather}
V_1(x)   =
V_0(x)-\frac{u''(x)}{u(x)}+\left[\frac{u'(x)}{u(x)}\right]^2
= - V_0(x) + 2\epsilon + \left[\frac{u'(x)}{u(x)}\right]^2 .
 \label{17}
\end{gather}
Hence, a straightforward calculation using equations
(\ref{22}), (\ref{upgenb0}) leads to:
\begin{gather}\label{23}
\left[\frac{u'(x)}{u(x)}\right]^2=k^2-\frac{4k^2(1-\tilde
a)}{u^2(x)} \big[D+\tilde aD^2H(-x)+\tilde aH(x)\big] .
\end{gather}
As $\epsilon = -k^2/2$, equations (\ref{17}), (\ref{23}) give:
\begin{gather}\label{24}
V_1(x)=a\delta(x)-\frac{4k^2(1-\tilde a)[D+\tilde aD^2H(-x)+\tilde
aH(x)]}{\{ e^{kx} + D e^{-kx} + 2\tilde a\sinh(kx)[DH(-x) -
H(x)]\}^2} .
\end{gather}
Note that the denominator of equation (\ref{24}) never vanishes for
$x\in(-\infty,\infty)$ and $D \geq 0$. Moreover, it can be seen that
the delta term in $V_1(x)$ is now repulsive (since $a>0$).

A straightforward consequence of the intertwining relationship
(\ref{1interwining}) is that for any eigenfunction $\psi$ of $H_0$
associated to the eigenvalue $E$ ($H_0\psi = E\psi$) such that
$A_1^+\psi \neq 0$, it turns out that $\psi^{(1)}\propto A_1^+\psi
\propto W(u,\psi)/u$ is a corresponding eigenfunction of $H_1$
associated to $E$. Moreover, if~$\psi$ satisf\/ies as well equation
(\ref{2}) it turns out that $\psi^{(1)}$ now obeys:
\begin{gather*}
\left(
 \begin{array}{c}
    \psi^{(1)}(0+) \\
 {\psi^{(1)}}'(0+)
               \end{array}
             \right)  =  \left(
      \begin{array}{cc}
        1 & 0 \\
        2a & 1
      \end{array}
    \right) \left(
               \begin{array}{c}
    \psi^{(1)}(0-) \\
 {\psi^{(1)}}'(0-)
               \end{array}
             \right) ,
\end{gather*}
which is consistent with the fact that the intensity of the delta
term in $V_1(x)$ has an opposite sign compared with $V_0(x)$ and the
second term of $V_1(x)$ has just a f\/inite discontinuity at $x=0$
(see equation (\ref{24})).

Concerning the spectrum of $H_1$, let us note in the f\/irst place
that $A_1^+$ transforms the scattering eigenfunctions of $H_0$ into
the corresponding ones of $H_1$. In particular, the wavefunction
$\psi(x)$ given in equation (\ref{ss}), when transformed by acting
on it with $A_1^+$, produces an expression $\psi^{(1)}(x)$ which is
a bit large to be presented here. However, for large values of
$\vert x\vert$ that expression reduces to the following scattering
one (up to a constant factor):
\begin{gather*}
   \psi^{(1)}(x)
\underset{\vert x\vert\rightarrow \infty}{\rightarrow}
\left[e^{i \kappa x} + \left(\frac{a}{a + i\kappa}\right)
\left(\frac{-k + i \kappa}{k + i \kappa}\right) e^{-i \kappa
x}\right] H(-x) \nonumber\\
\hphantom{\psi^{(1)}(x)
\underset{\vert x\vert\rightarrow \infty}{\rightarrow} }{}
 + \left(\frac{i\kappa}{a+i\kappa}\right)
\left(\frac{-k + i\kappa}{k + i\kappa}\right) e^{i \kappa x} H(x).
\end{gather*}
This means that the initial ref\/lection and transmission coef\/f\/icients
are unchanged under the f\/irst-order SUSY transformation (compare
equation (\ref{reftracoe})). We thus conclude that the continuous
spectrum of $H_0$ belongs as well to the spectrum of $H_1$.

Let us note that the dif\/ferences in the spectra of $H_1$ and $H_0$
rely in general in the modif\/ications produced by a non-singular SUSY
transformation on the discrete part of the initial spectrum. For
f\/irst-order transformations, these changes can be classif\/ied
according to the essentially dif\/ferent combinations of the
parameters $D\geq 0$ and $k\geq k_0$ which characterize the seed
eigenfunction~$u(x)$. We can f\/ind three dif\/ferent situations.

\begin{enumerate}\itemsep=0pt
\item[(i)] {\it Creation of a new ground state at $\epsilon <
E_0$.} This case appears for $D>0$, $k>k_0$. Here, the eigenfunction
$\psi_\epsilon^{(1)}\propto 1/u(x)$ of $H_1$ associated to
$\epsilon$ is square-integrable. Moreover, since the mapped initial
ground state $\psi_0^{(1)}
=\frac{1}{\sqrt{2}}\frac{1}{\sqrt{E_0-\epsilon}}
\frac{W(u,\psi_0)}{u}$ is as well a normalized eigenfunction of
$H_1$ with eigenvalue $E_0$, then Sp$(H_1) = \{\epsilon, E_0\}\cup
[0,\infty) = \{\epsilon\}\cup {\rm Sp}(H_0)$.

\item[(ii)] {\it Isospectral transformations.} These are achieved
from the previous case either by taking $D\rightarrow 0$ or
$D\rightarrow \infty$. Since in both situations $u(x)$ goes to zero
at one of the ends of the $x$-domain, it turns out that
$\psi_\epsilon^{(1)}\propto 1/u(x)$ is no longer square-integrable,
although $\psi_0^{(1)}$ is. Thus, Sp$(H_1) = \{E_0\}\cup [0,\infty)
= {\rm Sp}(H_0)$.

\item[(iii)] {\it Deleting $E_0$.} This situation arises from the
previous one by taking $k\!=\!k_0\!=\!a$ \mbox{($\tilde a \!=\! a/k \!=\! 1$)}. Since $u(x)
\propto \psi_0(x)$ is square-integrable, then
$\psi_\epsilon^{(1)}\propto 1/u(x)$ is not normalizable, and then
Sp$(H_1) = [0,\infty)$. From equation (\ref{24}), it is clear that
now
\begin{gather}\label{deltarespulsva}
V_1(x)=a\delta(x) .
\end{gather}
This means that, by deleting the bound state of the attractive delta
well $V_0(x) = -a\delta(x)$, $a>0$, which is placed at $E_0 = -a^2/2$,
we recover the repulsive delta barrier of equa\-tion~(\ref{deltarespulsva}), a standard result well known in the
literature.
\end{enumerate}

\section{Second-order SUSY transformations}\label{section4}

In this section it will be illustrated, by means of the delta-well
potential, the advantages for manipulating spectra of the
second-order SUSY transformations \cite{ais93,aicd95,bs97,fe97}
compared with the f\/irst-order ones. It is nowadays known that the
second-order SUSY partners $H_2$ of the initial Hamiltonian $H_0$
can be generated either by employing two eigenfunctions $u_1(x)$,
$u_2(x)$ of $H_0$, not necessarily physical, associated to two
dif\/ferent factorization energies $\epsilon_{1,2}$, $\epsilon_1 \neq
\epsilon_2$~\cite{ff05,fe10} or by an appropriate eigenfunction
$u_1(x)$ in the limit when $\epsilon_2\rightarrow \epsilon_1$ (the
so called conf\/luent case~\cite{mnr00,fs03}). In both situations the
two Hamiltonians $H_0$, $H_2$ are intertwined by a~second-order
operator in the way
\begin{gather*}
H_2 B_2^+ = B_2^+ H_0,
\end{gather*}
where
\begin{gather*}
B_2^+ = \frac12 \left(-\frac{d}{dx} +
\frac{{u_2^{(1)}}'}{u_2^{(1)}}\right) \left( -\frac{d}{dx} +
\frac{u_1'}{u_1} \right), \qquad u_2^{(1)} = \frac{w(x)}{u_1(x)},
\end{gather*}
the new Hamiltonian $H_2$ takes the standard Schr\"odinger form
\begin{gather*}
H_2 = -\frac12\frac{d^2}{dx^2} + V_2(x) ,
\end{gather*}
and the second-order SUSY partner $V_2(x)$ of the initial potential
$V_0(x)$ is given by
\begin{gather}\label{V2susy}
V_2(x) = V_0(x) - [\ln w(x)]'' ,
\end{gather}
the real function $w(x)$ being proportional in general to the
Wronskian of two generalized eigenfunctions of $H_0$~\cite{fs05}. An
explicit classif\/ication of the several second-order SUSY
transformations is next given.

\subsection{Conf\/luent case \cite{mnr00,fs03}}

Let us consider in the f\/irst place the limit $\epsilon_2 \rightarrow
\epsilon_1 \equiv \epsilon = - k^2/2 < 0$, taking as seed the
Schr\"odinger solution $u_+(x)$ vanishing as $x\longmapsto-\infty$,
which means to take the $u(x)$ given in equation (\ref{22}) with
$D=0$, namely:
\begin{gather}\label{25}
u(x) = e^{kx} - 2\tilde a\sinh(kx)H(x) .
\end{gather}
In this case the real function $w(x)$ appearing in equation
(\ref{V2susy}) takes the form \cite{fs03}
\begin{gather*}
w(x):=w_0+\int_{-\infty}^xu^2(y)\,dy .
\end{gather*}
An explicit calculation for $x\le 0$ leads to:
\begin{gather*}
\int_{-\infty}^xu^2(y)\,dy=\int_{-\infty}^x e^{2ky}\,dy =
\frac{e^{2kx}}{2k}.
\end{gather*}
On the other hand, for $x > 0$ it turns out that:
\begin{gather*}
\int_{-\infty}^xu^2(y)\,dy    =
\int_{-\infty}^0u^2(y)\,dy+\int_0^xu^2(y)\,dy\nonumber
\\
\hphantom{\int_{-\infty}^xu^2(y)\,dy }{}  =   \frac{\tilde a}{k} + \frac{(1 - 2\tilde a)}{2k}\,e^{2kx} + \frac{\tilde a^2}{k}\,\sinh(2kx)
+ 2\tilde a(1 - \tilde a)x  . 
\end{gather*}
By combining these two results, we obtain
\begin{gather}\label{29}
w(x) = w_0 + \frac{e^{2kx}}{2k} + \left[\frac{\tilde a}{k} -
\frac{\tilde a}{k} e^{2kx} + \frac{\tilde a^2}{k} \sinh (2kx) +
2\tilde a(1 - \tilde a)x \right] H(x) .
\end{gather}
The second-order SUSY partner potential of $V_0(x)$ becomes now
\begin{gather}
V_2(x)  =   - a \delta(x) + \frac{u^4(x)}{w^2(x)}  -
\frac{2u(x)u'(x)}{w(x)} .\label{30}
\end{gather}
Note that, since $u'(0+)=k-2a$ and $u'(0-)=k$, then $u'(x)$ and
consequently the potential dif\/ference $\Delta V(x) = V_2(x) -V_0(x)$
have a f\/inite discontinuity at $x=0$.

In order to avoid the arising of extra singularities for $V_2(x)$
with respect to $V_0(x)$ we have to take $w_0\geq 0$. Concerning the
spectrum of $H_2$, a similar calculation as in the f\/irst-order case
shows that the scattering eigenfunctions of $H_0$ are mapped into
the corresponding ones of $H_2$, i.e., the energy interval
$[0,\infty)$ belongs to Sp$(H_2)$. As for the discrete part of the
spectrum, several possibilities of spectral manipulation emerge,
according to how we choose $k$ and $w_0$.

\begin{figure}[t]
\centering
\includegraphics[width=9cm]{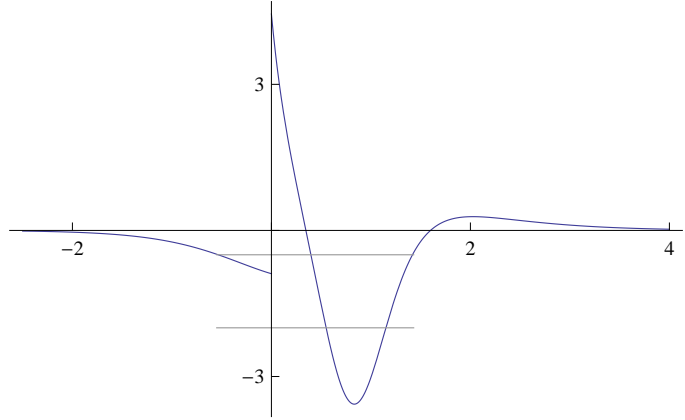}

\caption{Potential dif\/ference $\Delta V(x)$ as function of $x$ (blue
lines) induced by the conf\/luent second order SUSY transformation for
$a=2$, $k=1$, $w_0=1$. A new level was created at $\epsilon = -1/2$,
above the initial ground state $E_0=-2$ (gray horizontal lines).}
\label{fig:1}
\end{figure}

\begin{enumerate}\itemsep=0pt
\item[(i)] {\it Creating a new bound state at $\epsilon\neq
E_0$.} This case appears by taking $w_0>0$ and $k\neq k_0 = a$.
Since
\begin{gather*}
\lim_{\vert x\vert \rightarrow \infty} \psi_\epsilon^{(2)} \propto
\lim_{\vert x\vert \rightarrow \infty} \frac{u(x)}{w(x)} \propto
e^{-k\vert x\vert},
\end{gather*}
the eigenfunction $\psi_\epsilon^{(2)}$  of $H_2$ associated to
$\epsilon$ is square-integrable, i.e., a new bound state has been
created at $\epsilon$, either below the ground state for $k>k_0$ or
above it for $k<k_0$. The last option is illustrated in Fig.~\ref{fig:1},
where we have plotted the potential dif\/ference $\Delta V(x)$ as a
function of~$x$ for $a=2$, $k=1$, $w_0 = 1$, i.e., a new level was
created at $\epsilon = -1/2 > E_0 = -2$ (see the two gray horizontal
lines in the same graph). Note the existence of a f\/inite
discontinuity in $\Delta V(x)$ at $x=0$, induced by a similar
discontinuity of~$u'(x)$ at the same point.

\item[(ii)] {\it Isospectral transformations.} They arise in the
f\/irst place as a limit of the previous case for $\epsilon\neq E_0$
and $w_0\rightarrow 0$. Note that the long explicit expression for
the $V_2(x)$ of~(\ref{30}) which would appear if we would substitute
explicitly the~$u(x)$ and~$w(x)$ of equations (\ref{25}), (\ref{29})
becomes strongly simplif\/ied in this limit:
\begin{gather*}
  V_2(x) = - a \delta(x) + H(x) \Big(\!8\tilde a k^2 e^{2kx}[(\tilde
a-1)e^{2kx} - \tilde a]\{(\tilde a-1)[2kx(\tilde a - 1) + 1 -
2\tilde a]e^{2kx} \\
\hphantom{V_2(x) =}{} + \tilde a[2kx(\tilde a-1) + 2\tilde a -
3]\}\!\Big)\!\Big/ \! \big\{(\tilde a-1)^2e^{4kx} \! + 2\tilde a[1 - 2kx(\tilde a -
1)]e^{2kx}\! - \tilde a^2\big\}^2 .
\end{gather*}
Since now
\begin{gather*}
\lim_{x\rightarrow -\infty} \frac{u(x)}{w(x)} = \infty,
\end{gather*}
it turns out that $\epsilon\not\in{\rm Sp}(H_2) = \{E_0\}\cup
[0,\infty) = {\rm Sp}(H_0)$.

An alternative way to produce isospectral transformations is to use
the single bound state of $H_0$ for evaluating $w(x)$. The
corresponding formula is achieved from equation (\ref{29}) by taking
$k=k_0=a$, $\tilde a = 1$, which leads to:
\begin{gather}\label{it}
w(x) = w_0 + \frac{e^{2ax}}{2a} - \frac{2}{a}\sinh^2(ax)H(x)\,.
\end{gather}
Hence
\begin{gather}\label{isogro}
  V_2(x) = - a \delta(x) - \frac{8 w_0 a^3 e^{2ax}}{(2w_0a +
e^{2ax})^2}H(-x) + \frac{8 a^2 e^{2ax}(1+w_0a)}{[2(1 + w_0 a)e^{2ax}
- 1]^2}H(x) .
\end{gather}
Note that now $w(x)$ does not have any node for
\begin{gather*}
w_0 \in \left(-\infty, - \frac1a\right) \cup \left(0,\infty\right).
\end{gather*}
Moreover, in this domain it turns out that
\begin{gather*}
\lim_{\vert x\vert \rightarrow \infty} \frac{u(x)}{w(x)} \propto
e^{-a\vert x\vert},
\end{gather*}
i.e., $\psi_\epsilon^{(2)} \propto u(x)/w(x)$ is square-integrable
$\Rightarrow$ ${\rm Sp}(H_2) = \{E_0\}\cup [0,\infty) = {\rm
Sp}(H_0)$.

\item[(iii)] {\it Deleting the ground state of $H_0$.} By taking
now the limit of equation (\ref{it}) for $w_0 \rightarrow 0$ or $w_0
\rightarrow -1/a$, it turns out that $\lim\limits_{x\rightarrow -\infty}
u(x)/w(x) = \infty$ or $\lim\limits_{x\rightarrow \infty} u(x)/w(x) =
\infty$ respectively. In both cases $\psi_\epsilon^{(2)}$ is not
square-integrable and then
\begin{gather*}
E_0\not\in{\rm Sp}(H_2) = [0,\infty).
\end{gather*}
This result means that we have deleted the ground state of $H_0$ in
order to obtain $H_2$. For $w_0\rightarrow 0$ the potential of
equation (\ref{isogro}) becomes
\begin{gather}\label{delgro}
  V_2(x) = - a \delta(x) + \frac{8 a^2 e^{2ax}}{(2e^{2ax} -
1)^2}H(x) .
\end{gather}
On the other hand, for $w_0\rightarrow -1/a$ the corresponding
potential $V_2(x)$ is obtained from the previous one by the change
$x\rightarrow -x$.
\end{enumerate}

Let us remark that, although the f\/inal spectra of the SUSY partner
Hamiltonians of $H_0$ are the same when deleting its ground state in
the f\/irst-order and in the conf\/luent second-order transformations,
however the potentials $V_1(x)$ and $V_2(x)$ are physically
dif\/ferent (compare equations~(\ref{deltarespulsva}) and~(\ref{delgro})). In particular, note the opposite signs of the
coef\/f\/icients of the Dirac delta function for both potentials.

\subsection{Complex case \cite{fmr03,rm03,fr08}}

\begin{figure}[t]
\centering
\includegraphics[width=9cm]{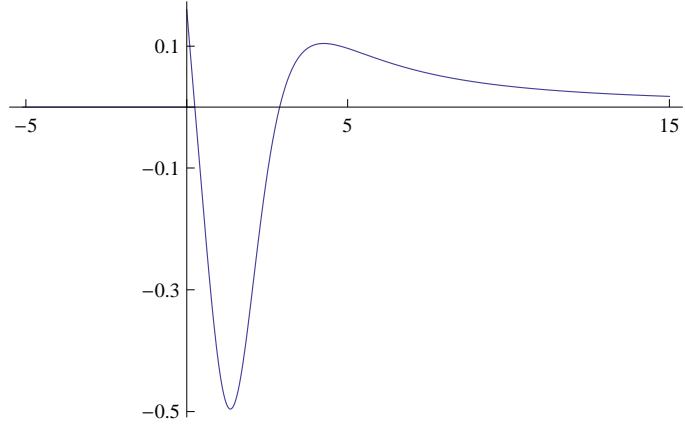}

\caption{Potential dif\/ference $\Delta V(x)$ as function of $x$
induced by the complex second order SUSY transformation for $a=2$, $k=1/100 + i/10$. The two potentials $V_2(x)$ and $V_0(x)$ are
isospectral.}
\label{fig:2}
\end{figure}

Let us assume that $k = k_R + i k_I$ is complex with $k_R>0$, $k_I
\in {\mathbb R}$, and suppose that the two involved factorization
energies are now given by $\epsilon = -k^2/2$ and $\bar \epsilon$,
where $\bar z$ denotes the complex conjugate of~$z$. Since we need
to avoid the arising of extra singularities in the new potential, we
will take a Schr\"odinger seed solution vanishing at one of the ends
of the $x$-domain in the form given in equation~(\ref{25}) with
$k\in{\mathbb C}$, namely,
\begin{gather}\label{31}
    u(x)=e^{kx}-\frac{2a}{k}\sinh(kx)H(x) ,\qquad
    \bar u(x)=e^{\bar kx} - \frac{2a}{\bar k}\sinh(\bar kx)H(x) .
\end{gather}
To compute now the second-order SUSY partner potential $V_2(x)$, we
have to obtain in the f\/irst place the Wronskian $W(u,\bar u)$ and
then the real function
\begin{gather*}
w(x)=\frac{W(u,\bar u)}{2(\epsilon - \bar \epsilon)}  .
\end{gather*}
This calculation is cumbersome but otherwise straightforward, which
leads to:
\begin{gather}
w(x)   =   \frac{e^{2k_Rx}}{2k_R} + \bigg\{ - \frac{a}{\vert
k\vert^2}\big[\cosh(2k_Rx)-\cos(2k_Ix)\big]
\nonumber\\
\phantom{w(x)   =}{} +\frac{a(a-k_R)}{\vert k\vert^2 k_Rk_I} \big[k_I\sinh(2k_Rx)
-k_R\sin(2k_Ix)\big]\bigg\} H(x) .\label{33}
\end{gather}
Then, $V_2(x)$ will be given by
\begin{gather}\label{v2complexcase}
V_2(x) = - a \delta(x) + \frac{\vert u(x)\vert^4}{w^2(x)} -
\frac{[u(x)\bar u'(x) + \bar u(x) u'(x)]}{w(x)},
\end{gather}
with $u(x)$, $\bar u(x)$ and $w(x)$ as given in equations (\ref{31})
and (\ref{33}). An illustration of the potential dif\/ference $\Delta
V(x)$ as function of $x$ for $a = 2$, $k = 1/100 + i/10$ is given in
Fig.~\ref{fig:2}.

Note that these equations become highly simplif\/ied if $k_R=a$:
\begin{gather}
w(x)   =   \frac{e^{2ax}}{2a} - \left( \frac{a}{a^2 + k_I^2} \right)\left[\cosh(2ax) - \cos(2k_Ix)\right]H(x) ,\label{34} \\
|u(x)|^2  =   e^{2ax} - \left( \frac{2a}{a^2 + k_I^2}\right)
\left[a\sinh(2ax)+k_I\sin(2k_Ix)\right]H(x) .\label{35}
\end{gather}
Moreover, for the particular choice $k_R=a$ we get a more compact
expression for the new potential~$V_2(x)$ than for a generic~$k_R$
that would appear if we would substitute the~$u(x)$ and~$w(x)$ of
equations~(\ref{31}),~(\ref{33}) in equation~(\ref{v2complexcase}):
\begin{gather}
  V_2(x) = - a \delta(x) + H(x)\Big(\! 4a^2
e^{2ax}[2(a^2-k_I^2)\cos(2k_Ix)(a^2-k_I^2 e^{4ax})\label{v2crg}\\
\hphantom{V_2(x) =}{}
-4ak_I\sin(2k_Ix)(a^2+k_I^2 e^{4ax}) + 8a^2k_I^2
e^{2ax}]\!\Big)\!\Big/\! \big[a^2-k_I^2 e^{4ax} - 2a^2 e^{2ax}\cos(2k_Ix)\big]^2 .\nonumber
\end{gather}

Let us remark that, for the general case characterized by equations
(\ref{31}) and (\ref{33}) as well as the particular ones described
by equations (\ref{34})--(\ref{v2crg}), the scattering states of $H_0$
are mapped into the corresponding ones of $H_2$, and the same
happens for the bound state. Thus, it turns out that the spectrum of
$H_2$ will be equal to ${\rm Sp}(H_0)= \{E_0\}\cup [0,\infty)$,
i.e., the complex second-order SUSY transformations which produce a
real f\/inal potential are strictly isospectral.

\subsection{Real case}

Let us take now two seed solutions $u_1$, $u_2$ in the form given in
equation (\ref{22}), associated to the pair of real factorization
energies $\epsilon_2 < \epsilon_1 \Rightarrow k_2 > k_1$. Their
explicit forms, and the corresponding derivatives, are given by:
\begin{gather}
u_i(x)       =   e^{k_i x} + D_i e^{- k_i x} + 2 \tilde
a_i\sinh(k_i x) [D_i H(-x) - H(x)]   , \label{seedsrc}
\\
u_i'(x)   =   - k_i u_i(x) + 2 k_i e^{k_i x}\left[1 - \tilde a_i
H(x) + D_i\tilde a_i H(-x) \right], \qquad i=1,2, \nonumber
\end{gather}
where $\tilde a_i = a/k_i$. Similarly as in the complex case, the
calculation of the Wronskian $w(x) \equiv W(u_1,u_2)$ of the two
involved Schr\"odinger seed solutions is once again cumbersome, but
a~convenient compact expression reads:
\begin{gather*}
w(x)       =     (k_1 - k_2) u_1 u_2 + 2 k_2 u_1 e^{k_2 x}\left[1 -
\tilde a_2 H(x) +
D_2 \tilde a_2 H(-x) \right] \nonumber \\
 \phantom{w(x)       =}{}  - 2 k_1 u_2 e^{k_1 x}\left[1 - \tilde a_1 H(x) + D_1 \tilde a_1
H(-x) \right]. 
\end{gather*}
By employing this equation, it is straightforward to calculate the
new potential through:
\begin{gather*}
V_2(x)       =      - a \delta(x) + \left( \frac{w'}{w} \right)^2 - \frac{w''}{w}     =   - a \delta(x) + \left[ \frac{(k_1^2 - k_2^2) u_1 u_2}{w}
\right]^2 + \frac{(k_1^2 - k_2^2) (u_1 u_2' + u_1' u_2)}{w} .
\end{gather*}

Concerning the spectrum of~$H_2$, once again the scattering states
of~$H_0$ are mapped into the corresponding ones of~$H_2$. As for the
discrete part of the spectrum, several possibilities are worth of
study.

\begin{enumerate}\itemsep=0pt

\item[(i)] {\it Creating two new levels.} Let us suppose f\/irst
that $\epsilon_1 \neq E_0 \neq \epsilon_2$. In order that $w(x)$  do not have nodes, the two factorization energies must be placed either both
below (for $k_2 > k_1 > a$) or both above $E_0$ (for $k_1 < k_2 <
a$). Moreover, according to the chosen ordering $\epsilon_2 <
\epsilon_1$, the solution $u_2(x)$ must have one extra node with
respect to $u_1(x)$ \cite{ff05}. In the domain $k_2 > k_1 > a$
($\epsilon_2 < \epsilon_1 < E_0$) this can be achieved by taking
$D_2 < 0$ and $D_1 > 0$ while for $k_1 < k_2 < a$ ($E_0 < \epsilon_2
< \epsilon_1$) it must be taken $D_2 > 0$ and $D_1 < 0$. With this
choice of parameters, it turns out that the two eigenfunctions of
$H_2$ associated to $\epsilon_1$ and $\epsilon_2$,
$\psi_{\epsilon_1}^{(2)} \propto u_2/w$ and $\psi_{\epsilon_2}^{(2)}
\propto u_1/w$, are square-integrable. Thus,
\begin{gather*}
{\rm Sp}(H_2) = \{\epsilon_2,\epsilon_1\}\cup {\rm Sp}(H_0),
\end{gather*}
i.e., two new levels have been created for $H_2$, either both below
the ground state of $H_0$ (for $k_2 > k_1 > a$) or both above $E_0$
(for $k_1 < k_2 < a$). An illustration of the last situation is
shown in Fig.~\ref{fig:3}, where we have plotted the potential dif\/ference
$\Delta V(x)$ for $a=2$, $k_2 = 1$, $k_1=1/2$, $D_1 = -1/2$, $D_2 =
1$. As a result of the transformation, two new levels were created
above the ground state energy of $H_0$ at the positions $\epsilon_2
= -1/2$ and $\epsilon_1 = -1/8$ (see the gray horizontal lines at
Fig.~\ref{fig:3}).

\begin{figure}[t]
\centering
\includegraphics[width=9cm]{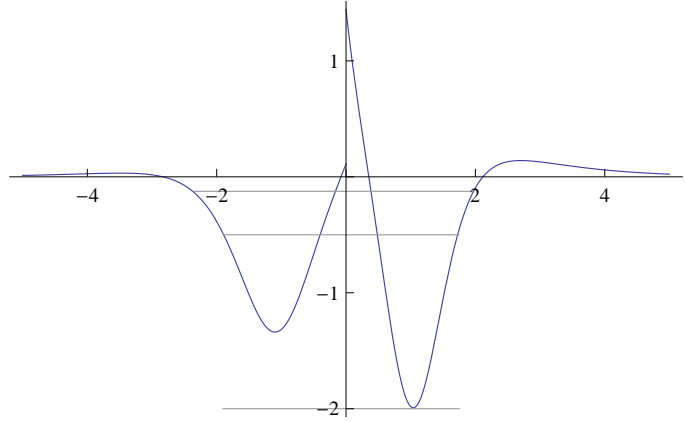}

\caption{Potential dif\/ference $\Delta V(x)$ as function of $x$ (blue
lines), induced by a real second order SUSY transformation for $a=2$,
$k_2 = 1$, $k_1=1/2$, $D_1 = -1/2$, $D_2 = 1$. Note that two new
levels were created above $E_0 = -2$, at the positions $\epsilon_2 =
-1/2$ and $\epsilon_1 = -1/8$ (gray horizontal lines).}
\label{fig:3}  
\end{figure}

\item[(ii)] {\it Creating one new level.} This case arises from
the previous one for $D_2 \rightarrow 0$. Now it turns out that
$\psi_{\epsilon_2}^{(2)}$ is not square-integrable anymore, meaning
that
\begin{gather*}
{\rm Sp}(H_2) = \{\epsilon_1\}\cup {\rm Sp}(H_0).
\end{gather*}
Thus, in order to generate $H_2$ a new level has been created at
$\epsilon_1$, above $E_0$ for $k_1 < a$ and below it for $k_1>a$.

\item[(iii)] {\it Isospectral transformations.} These can be
achieved from case (i) for $D_1 = D_2 \rightarrow 0$, where both
$\psi_{\epsilon_1}^{(2)}$ and $\psi_{\epsilon_2}^{(2)}$ cease to be
square-integrable so that $\epsilon_i \not\in{\rm Sp}(H_2)$, $i=1,2$. Hence,
\begin{gather*}
{\rm Sp}(H_2) = {\rm Sp}(H_0).
\end{gather*}

\item[(iv)] {\it Moving the level $E_0$.} This procedure is
obtained from case~(i), e.g., by taking $\epsilon_2 = E_0$, $D_2
\rightarrow 0$, $u_2(x) \propto \psi_0(x)$,  and $u_1(x)$ as given
in equation~(\ref{seedsrc}) with $D_1 < 0$, $\epsilon_1 > E_0$. With
this choice it can be shown that $\psi_{\epsilon_2}^{(2)} \propto
u_1/w$ is not square-integrable but $\psi_{\epsilon_1}^{(2)}$ does,
meaning that
\begin{gather*}
{\rm Sp}(H_2) = \{\epsilon_1\}\cup [0,\infty).
\end{gather*}
In a way, the level $E_0$ has been moved up to $\epsilon_1$ for
generating $H_2$.

\item[(v)] {\it Deleting the level $E_0$.} This can be achieved
as a limit of the previous case for $D_1 \rightarrow 0$. Now it
turns out that $\lim\limits_{x\rightarrow 0}u_2/w = \infty$, i.e.,
$\epsilon_1 \not\in{\rm Sp}(H_2)$, and hence
\begin{gather*}
{\rm Sp}(H_2) = [0,\infty).
\end{gather*}
\end{enumerate}

\section{Conclusions}\label{section5}

We have employed the f\/irst and second-order supersymmetric quantum
mechanics for generating new potentials with modif\/ied spectra
departing from the delta well potential. The f\/irst-order
transformation allowed us to change just the ground state energy
level, while the second-order transformations enlarged the
possibilities of spectral control, including the option of
manipula\-ting the excited state levels. On the other hand, it is
important to remember that the f\/irst-order transformations induced
in the new potential a delta term with an opposite sign compared
with the initial one (physically the delta term changed from
attractive to repulsive). Meanwhile, the second-order
transformations generated a delta term with exactly the same sign as
the initial one (the attractive nature was preserved under the
transformation). These physical dif\/ferences should be taken into
account in the determination of the most appropriate transformation
for building a potential model. We can conclude that supersymmetric
quantum mechanics is a~po\-wer\-ful mathematical tool, which is quite
useful for implementing the spectral design in physics.

\subsection*{Acknowledgement}

Partial f\/inancial support is acknowledged to the Spanish Junta de
Castilla y Le\'on (Project GR224) and the Ministry of Science and
Innovation (Projects MTM2009-10751 and FPA2008-04772-E). DJFC
acknowledges the support of Conacyt.

\pdfbookmark[1]{References}{ref}
\LastPageEnding

\end{document}